\documentclass[aps,hidelinks,twocolumn, prl, superscriptaddress,float fix]{revtex4-2}
\usepackage{graphicx}
\usepackage{orcidlink}
\usepackage{amsmath,amssymb,amsfonts}
\usepackage[normalem]{ulem}
\usepackage{graphicx}
\usepackage{dcolumn}
\usepackage{bm,xcolor}
\usepackage{hyperref}

\begin{document}

\preprint{APS/123-QED}

\title{Non-invertible circuit complexity from fusion operations}

\author{Saskia Demulder\,\orcidlink{0000-0002-4337-4489}}
\email{saskia.demulder@cunef.edu}
\affiliation{Department of Quantitative Methods, CUNEF Universidad,\\ Calle Almansa 101, 28040 Madrid, Spain}

\begin{abstract}
Modern understanding of symmetry in quantum field theory includes both invertible and non-invertible operations. Motivated by this, we extend Nielsen's geometric approach to quantum circuit complexity to incorporate non-invertible gates. These arise naturally from fusion of topological defects and allow transitions between superselection sectors. We realise fusion operations as completely positive, trace-preserving quantum channels. Including such gates makes the sector-changing optimisation problem discrete: it reduces to a weighted shortest-path problem on the fusion graph. Circuit complexity therefore combines continuous geometry within sectors with discrete sector jumps. We illustrate the framework in rational conformal field theories and briefly comment on an AdS$_3$ interpretation in which fusion-induced transitions correspond to geometry-changing boundary operations. A companion paper provides full derivations and extended examples.
\end{abstract}

\maketitle


\section{\label{sec:level1}Introduction}
Quantum circuit complexity provides a geometric framework for quantifying the resources required to implement transformations using a prescribed set of elementary gates, and has played an increasingly central role in connecting quantum information, quantum field theory, and gravity \cite{Baiguera:2025dkc,Chapman:2021jbh}. In Nielsen's formulation \cite{Nielsen:2005mkt,Nielsen:2006cea}, one considers continuous families of invertible gates, which generate trajectories on a group manifold; in quantum many-body and field-theoretic applications these gates are naturally identified with symmetry transformations acting within a fixed superselection sector. While powerful, this structure implies a basic limitation: such circuits cannot change superselection sectors or, in two-dimensional conformal field theories, move between distinct conformal families.

From a modern viewpoint, this limitation is not fundamental. Symmetries in quantum field theory are now understood more broadly as topological operators rather than group actions \cite{Gaiotto:2014kfa,Bhardwaj:2023kri,Schafer-Nameki:2023jdn}. In this setting, many theories admit intrinsically non-invertible symmetries governed by fusion, realised through topological defects and categorical symmetry operators across high-energy and condensed-matter systems \cite{Petkova:2000ip,Fuchs:2002cm,Frohlich:2006ch,McGreevy:2022oyu,Shao:2023gho}. Acting with such a symmetry generally produces multiple admissible outcomes associated with different superselection sectors, rendering the operation incompatible with a purely unitary gate acting on a fixed Hilbert space.

In this work we show that non-invertible symmetries can nevertheless be incorporated consistently into a circuit-complexity framework by realising fusion as a completely positive, trace-preserving quantum channel. Using a Stinespring dilation, fusion gates are realised as reversible embeddings into an enlarged Hilbert space, with associativity controlled by the fusion associator of an underlying unitary modular tensor category \cite{bakalov2001lectures}. Circuits built from such gates combine continuous invertible evolution with discrete, sector-changing steps, and optimisation of the sector-changing part reduces to a shortest-path problem on the fusion graph of superselection sectors. Related perspectives on non-invertible symmetries as quantum operations have appeared recently in the QFT literature \cite{Okada:2024qmk}.

We illustrate the construction in rational conformal field theories \cite{Jefferson:2017sdb,Chagnet:2021uvi,Erdmenger:2020sup,Susskind:2018pmk} and offer an AdS$_3$-motivated interpretation \cite{Caputa:2018kdj,Erdmenger:2021wzc,Erdmenger:2022lov} in which fusion-induced transitions correspond to discrete jumps in boundary stress-tensor data. More generally, the framework applies to quantum systems with intrinsic superselection structure, where symmetry actions are generically non-invertible and organised by fusion rather than group multiplication, as occurs for categorical and duality symmetries in quantum field theory and topological phases of matter. This implies that circuit complexity in such systems is generically not captured by purely unitary, group-manifold-based geometric frameworks, but instead acquires an unavoidable discrete component associated with sector-changing operations.

\section{Fusion as a Non-Invertible Circuit Element}

\subsection{Fusion as a sector-changing operation}
In rational two-dimensional conformal field theories, fusion defines a fundamental operation that combines superselection sectors \cite{Petkova:2000ip,Fuchs:2002cm}. 
Throughout this section, we use labels $a,c$ to denote superselection sectors of quantum states (with Hilbert spaces $\mathcal H_a$), while a label $b$ denotes a fusion operation or topological defect acting on those states. For simple objects $a$ and $b$, fusion is encoded by the fusion rules
\begin{align}\label{eq:fusion}
a\otimes b =\bigoplus_c N_{ab}^{c}\,c\,,
\end{align}
where the sum runs over admissible output sectors $c$ with finite multiplicities $N_{ab}^{\,c}$.  For each allowed outcome $c$, the multiplicity $N_{ab}^{\,c}$ labels independent fusion channels, commonly referred to as junctions or intertwiners.  As an operation, fusion has a defining feature: it admits multiple outputs and therefore does not possess an inverse.

From a circuit perspective, this already marks a sharp departure from standard gate operations. 
A unitary gate acts invertibly within a fixed Hilbert space, whereas fusion maps states supported on a single sector $a$ to states supported on several distinct sectors $c$. 
No operation acting on the physical Hilbert space alone can reconstruct the original sector from a given output branch. 
This lack of invertibility is not a pathology, but a direct consequence of the superselection structure intrinsic to quantum field theory.

Importantly, the resulting non-unitarity is not a statement about loss of norm or breakdown of quantum mechanics, but reflects the loss of information about which fusion channel occurred. Fusion is therefore a symmetry operation whose action cannot be represented as a unitary gate acting on a fixed Hilbert space, but necessarily involves discarding sector-resolving data.

\subsection{Fusion as a quantum channel} 
To incorporate fusion into a circuit framework, we realise fusion as a quantum channel \cite{NielsenChuang,PreskillQI}. Fixing an input sector $a$ and a defect label $b$, fusion with $b$ defines a completely positive, trace-preserving (CPTP) map acting on density operators supported on $\mathcal H_a$,
\begin{align}\label{eq:channel}
\mathcal E_b(\rho_a)= \sum_{c,\mu} K^{(a,b)}_{c,\mu}\,\rho_a\,K^{(a,b)\dagger}_{c,\mu}.
\end{align}
Here the Kraus operators $K^{(a,b)}_{c,\mu} : \mathcal H_a \to \mathcal H_c$ are labelled by the fusion channel $c$ and a degeneracy index $\mu=1,\dots,N_{ab}^{\,c}$. Each Kraus operator corresponds to an intertwiner implementing the fusion transition $a \to c$.

Trace preservation, 
\begin{align}\label{eq:pivot}
\sum_{c,\mu} K^{(a,b)\dagger}_{c,\mu} K^{(a,b)}_{c,\mu}= \mathbf 1_{\mathcal H_{a}}\,,
\end{align}
ensures that fusion maps normalised states to normalised states. This condition is guaranteed by the unitarity of the underlying modular tensor category, which endows fusion spaces with canonical Hermitian structures \cite{bakalov2001lectures}. While individual Kraus operators depend on a basis choice in the fusion spaces, the resulting quantum channel $\mathcal E_b$ is basis-independent.

\begin{figure}[t]
    \centering
    \includegraphics[width=0.82\linewidth]{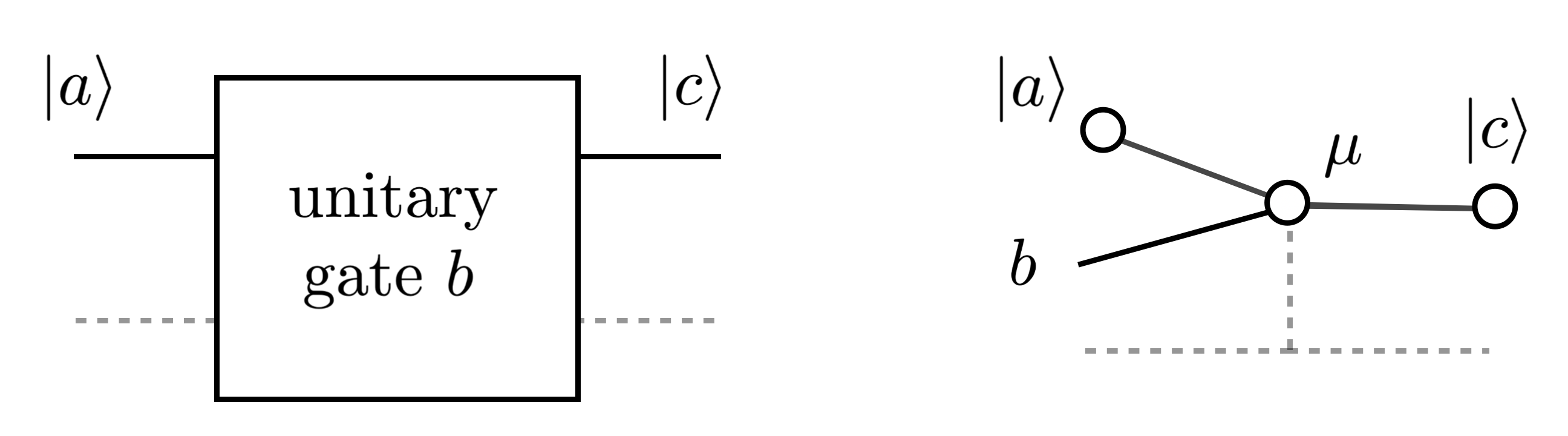}
    \caption{A unitary gate acts invertibly within a fixed superselection sector. By contrast, fusion with a defect $b$ defines a quantum channel whose Kraus operators $K^{(a,b)}_{c,\mu}$ correspond to junctions (intertwiners) mapping an input sector $a$ to admissible output sectors $c$. Retaining the channel label $\mu$ yields a reversible embedding into a larger Hilbert space, while discarding it produces a trace-preserving but non-unitary map on the physical Hilbert space.
}
    \label{fig:extending_VS}
\end{figure}

This formulation makes precise in what sense fusion is non-invertible. Retaining the fusion-channel label yields a reversible embedding (an isometry) into a larger Hilbert space, while discarding it produces a non-unitary but trace-preserving map on the physical Hilbert space (Fig. \ref{fig:extending_VS}). In this respect, fusion closely parallels a measurement process \cite{NielsenChuang,PreskillQI}: unitarity is restored only when the auxiliary degrees of freedom recording the outcome are retained.

Fusion gates therefore implement discrete transitions between superselection sectors. Circuits built from such gates are no longer described by smooth trajectories on a group manifold, but by sequences of admissible sector changes. This structure underlies the modified optimisation problem for circuit complexity discussed below.

\subsection{Consistency under composition}
A basic consistency requirement for any circuit construction is that successive gate applications define an unambiguous operation, independent of how they are grouped. For fusion gates, this is nontrivial: successive fusions admit different parenthesisations corresponding to distinct fusion trees, which a priori could define different quantum channels.

At the level of fusion rules, associativity is ensured by the structure of the underlying unitary modular tensor category. When fusion is realised as a quantum channel, this associativity persists in a precise operational sense. Consider the fusion of three defect labels $a,b,c$, whose total fusion may produce an output sector $y$ with nonzero multiplicity. The Kraus operators associated with different fusion orderings are related by unitary $F$-moves,
\begin{align}
K^{((ab)c)}_{y,\alpha} = \sum_{\beta} (F^{abc}_y)_{\alpha\beta}\, K^{(a(bc))}_{y,\beta}\,,
\end{align}
where $\alpha,\beta=1,\dots,N^{\,y}_{abc}$ label the degeneracy of the fusion channel leading to the output sector $y$. Since the $F$-symbols are unitary, the corresponding Kraus representations define the same completely positive trace-preserving map.

Fusion channels are therefore associative under composition, even though the individual fusion gates are non-invertible. At the same time, their non-invertible character implies that fusion gates cannot, in general, be freely reordered with invertible evolution. Full derivations and further discussion are given in the companion paper \cite{companion}.

\section{Circuit complexity with fusion gates}

Invertible gates correspond to unitary symmetry transformations acting within a fixed superselection sector. Their cost is measured exactly as in the standard Nielsen approach, by assigning a continuous cost functional to unitary circuit segments. Fusion gates, by contrast, change the superselection sector and are intrinsically non-invertible, introducing a qualitatively new contribution to circuit complexity.

A circuit implementing a transformation between an initial and a target state therefore consists of continuous unitary evolution within each sector, interspersed with fusion operations that jump between sectors. We assign the total cost additively,
\begin{align}
\mathcal C=\sum_{\mathrm{unitary\, segments}} \mathcal C_\mathrm{inv}+\sum_{\mathrm{fusion\, gates}}\mathcal C_\mathrm{fusion}\,,
\end{align}
where $\mathcal C_\mathrm{inv}$ is the standard Nielsen cost and $\mathcal C_\mathrm{fusion}$ is a positive cost associated with each admissible fusion operation. The precise functional form of $\mathcal C_\mathrm{fusion}$ does not affect the structural consequence discussed below, namely the reduction of circuit optimisation to a discrete shortest-path problem. A schematic construction is outlined in the End Matter, while a general formulation and extensions relevant for applications to $3d$ gravity are presented in the companion paper.

The inclusion of fusion gates fundamentally alters the optimisation problem. Any circuit connecting two superselection sectors must correspond to a finite sequence of admissible fusion steps, whose structure is fixed by the fusion rules. Assigning a cost to each such step endows the space of sectors with a weighted graph structure, where an edge between sectors $a$ and $c$ exists precisely when $N_{ab}^{\,c}\neq 0$ in Eq. \eqref{eq:fusion}.

\begin{figure}[t]
    \centering
    \includegraphics[width=0.72\linewidth]{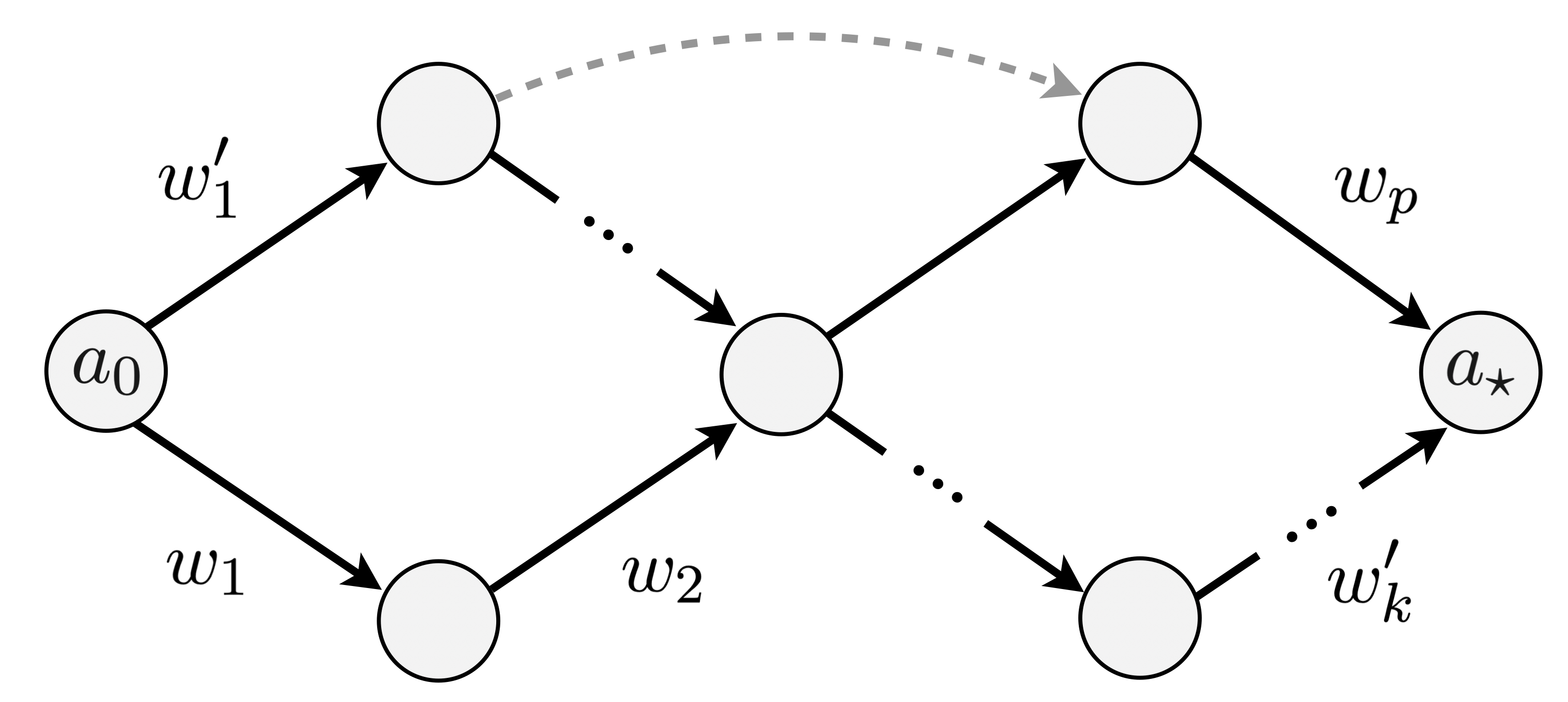}
    \caption{Schematic fusion graph of superselection sectors. Vertices denote sectors and directed edges correspond to admissible fusion operations, weighted by gate costs $w_i$. Distinct paths from an initial sector $a_0$ to a target sector $a_\star$ represent different fusion circuits with the same endpoints but different total costs. Circuit optimisation reduces to a shortest-path problem on this graph.}
    \label{fig:graph}
\end{figure}

Circuit optimisation therefore reduces to a shortest-path problem on this fusion graph (Fig. \ref{fig:graph}), with vertices labelled by sectors and edges weighted by fusion costs. Continuous Nielsen geometry governs invertible evolution within each vertex, while fusion gates implement transitions between vertices. Once fusion gates are assigned any nonzero cost, optimisation necessarily reduces to a shortest-path problem on the fusion graph, independent of the detailed cost functional.

\subsection*{Example: Ising fusion}
We illustrate the general framework using the Ising model, which provides the simplest non-trivial example of a non-invertible fusion rule. The Ising category contains three simple objects $\{\mathbf 1,\sigma,\varepsilon\}$, with fusion
$\sigma \times \sigma = \mathbf 1 \oplus \varepsilon$.
Together with $\varepsilon \times \sigma = \sigma$, this already captures the essential new features introduced by non-invertible fusion gates.

From a circuit perspective, fusion with $\sigma$ is intrinsically non-unitary: acting on a state in the $\sigma$ sector produces an output supported on two distinct superselection sectors $\mathbf 1$ and $\varepsilon$. As discussed above, this operation is therefore naturally realised as a quantum channel rather than a unitary gate.

Consider circuits that begin and end in the $\sigma$ sector but differ in their intermediate fusion structure. Such circuits necessarily involve two $\sigma$-fusion steps, with an intermediate channel that may be either $\mathbf 1$ or $\varepsilon$. While both choices lead to the same overall sector transition, they correspond to distinct fusion paths in the sense of the fusion graph. Once fusion gates are assigned a nonzero cost, these paths are no longer equivalent from the perspective of circuit optimisation.

This example illustrates two general features of fusion-based circuit complexity. First, non-invertible gates introduce genuine branching into circuit evolution, even in the simplest rational CFT. Second, circuit optimisation depends not only on the initial and final sectors but on the full fusion path connecting them. The Ising model thus provides a minimal setting in which the discrete, path-dependent nature of fusion-based circuit complexity becomes manifest. Extensive examples are discussed in the companion paper \cite{companion}.

\section{AdS\texorpdfstring{$_3$}{3} Interpretation and Outlook}

As one illustrative application, we briefly comment on an interpretation in two-dimensional conformal field theories that admit a semiclassical AdS$_3$ description \cite{Brown:1986nw}.  In such theories, superselection sectors are labelled by conformal families; restricting attention to rational subsectors, where the fusion data are finite \cite{Bershadsky:1989mf,Castro:2011zq,Janik:2025zji}, fusion operations shift the associated conformal weights $h_c$.

In the Chern-Simons formulation of AdS$_3$ gravity \cite{Achucarro:1986uwr,Witten:1988hc}, classical solutions are characterised by flat connections whose holonomy data are fixed by boundary stress-tensor profiles \cite{Banados:1998gg}. Changes in conformal weight therefore correspond to discrete changes in this boundary data. From this perspective, the action of a non-invertible fusion gate naturally induces a transition between distinct classical configurations, rather than a continuous deformation within a single solution. A more detailed discussion of the associated energy jumps and their interpretation
as geometry-changing circuit operations is given in the End Matter.

At the level of boundary stress-tensor data, fusion-induced shifts of conformal weight are
encoded as discrete jumps,
\begin{align}\label{eq:jump_bdy_tens}
    \Delta \mathcal L \sim \Delta h_c\,\delta(x^+)\,,\quad \Delta \bar{\mathcal L} \sim \Delta \bar h_c\,\delta(x^-)\,.
\end{align}
where $x^+$ is a chiral boundary coordinate, $\mathcal L$ denotes the Ba\~nados stress-tensor function, and $\Delta h_c$ the shift associated with the fusion outcome. This expression captures a change in boundary data specifying the classical solution, rather than a propagating bulk excitation. The initial and final Ba\~nados geometries are distinct classical solutions, corresponding to different Virasoro coadjoint orbits. In particular, fusion-induced energy shifts may interpolate between global AdS$_3$ or conical defect geometries and BTZ black hole geometries, rather than along deformations of a single geometry.

The circuit framework developed here is structural rather than dynamical and applies broadly to quantum systems with superselection structure. Two-dimensional conformal field theories with a semiclassical AdS$_3$ dual provide a particularly clean setting in which the geometric implications of fusion-induced sector changes can be made explicit. From this perspective, fusion gates connect sectors associated with different classical boundary data, rather than generating continuous motion along deformations of a fixed geometry. Fusion thus acts as a boundary operation that reorganises the classical description, producing localised energy-injection features in the boundary stress tensor.

\paragraph*{Discussion and outlook.}
Although we have illustrated this framework in rational subsectors of two-dimensional conformal field theory, where fusion data are finite and explicitly computable, the underlying construction is not restricted to this setting. More generally, whenever a quantum system possesses intrinsic superselection sectors organised by fusion, circuit complexity necessarily acquires a sector-changing component reflecting the algebraic structure of sector changes, rather than solely continuous unitary evolution within a fixed sector.

While fusion is also a central ingredient in topological quantum computation \cite{Nayak_2008}, its rôle here is conceptually distinct. In the present context, fusion acts as a non-invertible, sector-changing operation that structures circuit complexity itself, rather than as a logical gate acting within a protected computational subspace. This perspective suggests that non-invertible symmetries provide a natural organisational principle for circuit complexity in a wide class of quantum systems. In particular, analogous discrete contributions to complexity are expected to arise in higher-dimensional gravity theories and lattice systems with intrinsic non-invertible symmetries, where fusion-like sector changes are fundamental.

\vspace{10pt}

\begin{acknowledgments}
I would like to thank Stefano Baiguera, Shira \mbox{Chapman}, Rathindra N. Das and  Arnab Kundu for interesting discussions and comments. I would also like to thank  DESY and Universit\"at Hamburg, where part of this work was carried out, for the kind hospitality.
\end{acknowledgments}

\vspace{10pt}

\paragraph{End Matter on continuous versus discrete complexity in two-dimensional CFTs.}
In two-dimensional conformal field theories, invertible symmetry-based circuits generated by the Virasoro algebra act within a fixed highest-weight representation. In the Nielsen formulation, such circuits define continuous trajectories on a single Virasoro coadjoint orbit $\mathcal O_h$ \cite{Caputa:2018kdj}, and their complexity is governed by the Alekseev-Shatashvili geometric action, where $f(\sigma,\tau)$ parametrises the circuit trajectory. This yields a smooth geometric optimisation problem, and admits a direct AdS$_3$ interpretation when the instantaneous generator is identified with the physical CFT Hamiltonian \cite{Erdmenger:2021wzc}. 

Such circuits are intrinsically sector-preserving: the conformal weight $h$ labelling the coadjoint orbit is fixed. Non-invertible fusion operations instead induce transitions or jumps $\mathcal O_h \rightarrow \mathcal O_{h_c}$ corresponding to discrete jumps between inequivalent coadjoint orbits. A general circuit $D_{g_1}D_{a_1}D_{g_2}\cdots |0\rangle$ that encompasses both invertible $D_{g_i}$ and non-invertible gates $D_{a_i}$, therefore explores a stratified configuration space consisting of continuous orbits connected by  sector-changing transitions. At each transition induced by a non-invertible gate, the Alekseev-Shatashvili geometric action measure the invertible cost is updated. Circuit complexity acquires an intrinsically hybrid structure, with a continuous geometric contribution within each orbit and a discrete contribution associated with fusion-induced jumps, which cannot be captured by purely geometric metrics on a group manifold.

\vspace{10pt}
\paragraph{End Matter on  non-invertible gates and Stinespring dilation.}
Fusion with a defect labelled by $b$ acts between superselection sectors and is intrinsically non-unitary. Operationally, it is realised as a completely positive, trace-preserving map
\begin{align}
\mathcal E_b(\rho_a) = \sum_{c,\mu} K^{(a,b)}_{c,\mu}\,\rho_a\,K^{(a,b)\dagger}_{c,\mu}\,,
\end{align}
where the Kraus operators $K^{(a,b)}_{c,\mu}:\mathcal H_a\to\mathcal H_c$ are fusion intertwiners labelled by the fusion outcome $c$ and multiplicity index $\mu$. This channel arises from a canonical Stinespring isometry
\begin{align}
W_{ab}:\mathcal H_a \longrightarrow  \bigoplus_c \mathcal H_c\otimes V^c_{ab}\,,
\end{align}
with $V^c_{ab}$ the fusion spaces of the underlying unitary modular tensor category. The Stinespring isometry is the realisation of the a non-invertible gate $D_b$ acting on a state in the superselection sector $a$. 

The completeness relation
\begin{align}
\sum_{c,\mu}
K^{(a,b)\dagger}_{c,\mu}K^{(a,b)}_{c,\mu} = \mathbf 1_{\mathcal H_a}
\end{align}
follows from the pivotal structure and categorical trace identities. Different orderings of fusion gates yield unitarily equivalent Kraus representations, with equivalence implemented by the associator or $F$-symbols. Thus, categorical consistency replaces group structure: non-invertible gates compose consistently as quantum channels, even though they are not invertible operations. The associated optimisation problem is discrete and graph-like, rather than geodesic.

The quantum-channel realisation of fusion gates provides a natural starting point for defining costs associated with non-invertible operations. For a fixed fusion process $(a,b)$ with admissible outcomes $c$, the Stinespring construction selects canonical channel weights
\begin{align}
p(c|a,b)=\frac{N^{\,c}_{ab}\,d_c}{d_a d_b}\,,
\end{align}
determined solely by fusion multiplicities and quantum dimensions. These weights define a point in a finite-dimensional projective space of fusion outcomes, on which a basis-independent distance can be introduced. In particular, identifying the vector $(\sqrt{p(c|a,b)})_c$ with a ray, one obtains a natural Fubini-Study metric that assigns an intrinsic geometric cost to a fusion gate, measuring the deviation from unbiased channel branching. This intrinsic cost captures the kinematical structure of fusion-induced branching. Further refinements, including energy-sensitive weights and outcome-selection costs required for physical applications, follow directly from the quantum-channel interpretation and are developed in detail in the companion paper.

\vspace{10pt}
\paragraph{End Matter on AdS$_3$ interpretation with geometry-changing circuits.}
In two-dimensional conformal field theories for which an AdS$_3$ interpretation of the Virasoro data is available, superselection sectors are labelled by Virasoro highest weights $(h,\bar h)$, with boundary Hamiltonian
\begin{align}
H = L_0+\bar L_0-\frac{c}{12}\,,\quad E(h,\bar h)=h+\bar h-\frac{c}{12}\,.
\end{align}
Fusion-induced changes of conformal family therefore produce discrete boundary energy jumps,
\begin{align}
\Delta E=(h_c+\bar h_c)-(h+\bar h)\,.
\end{align}

In the Chern-Simons formulation of AdS$_3$ gravity, classical solutions are specified by flat $\mathrm{SL}(2,\mathbb R)\times\mathrm{SL}(2,\mathbb R)$ connections whose holonomy classes are fixed by constant boundary stress-tensor data through $T(x^+)=\tfrac{c}{12}\mathcal L$ and  $\bar T(x^-)=\tfrac{c}{12}\bar{\mathcal L}$.

Invertible Virasoro circuits generate continuous evolution within a fixed coadjoint orbit and, under the interpretation of \cite{Erdmenger:2021wzc} correspond to time evolution of a single bulk geometry. By contrast, fusion gates change $(h,\bar h)$ and therefore modify the holonomy class itself. At the level of boundary data, this may be represented schematically as in Eq. \eqref{eq:jump_bdy_tens} signalling an instantaneous change of geometry rather than propagating bulk dynamics.

Whenever the fusion outcome satisfies $h_c+\bar h_c>\frac{c}{12}$ the corresponding boundary data describe a BTZ black hole rather than global AdS$_3$ or a conical defect. Circuit complexity including fusion gates therefore measures the cost of geometry-changing operations: boundary processes that inject finite energy and reorganise the global holonomy data specifying the classical solution. This interpretation is purely kinematical: no real-time gravitational dynamics is assumed. Fusion gates act as boundary operations that modify the data specifying the bulk geometry, and circuit complexity characterises the minimal cost of such geometry-changing operations  within a boundary description.

\bibliography{biblio}

\end{document}